\documentstyle[12pt,prc,aps,floats,epsfig]{revtex}
\tighten
\begin{document}
\draft

\title{Accurate Nucleon-Nucleon Potential Based upon Chiral Perturbation
      Theory}

\author{D. R. Entem\footnote{On leave from University of Salamanca, Spain.
        E-mail: dentem@uidaho.edu}
        and R. Machleidt\footnote{E-mail: machleid@uidaho.edu}}

\address{Department of Physics, University of Idaho, Moscow, ID 83844, USA}

\date{\today}

\maketitle

\begin{abstract}
We present an accurate nucleon-nucleon ($NN$) potential
based upon chiral effective Lagrangians.
The model includes one- and two-pion exchange contributions
up to chiral order three. We show that a quantitative fit
of the $NN$ $D$-wave phase shifts requires contact terms (which represent
the short range force) of order four.
Within this framework, the $NN$ phase shifts below 300 MeV lab.\
energy and the properties of the deuteron
are reproduced with high-precision.
This chiral $NN$ potential represents a reliable starting point
for testing the chiral effective field theory approach in exact
few-nucleon and microscopic nuclear many-body calculations.
An important implication of the present work is that the chiral $2\pi$ exchange
at order four is of crucial interest for future chiral $NN$
potential development.
\end{abstract}

\vspace*{1cm}

One of the most fundamental problems of nuclear physics is to
derive the force between two nucleons from first principles.
A great obstacle for the solution of this problem has been the fact
that the fundamental theory of strong interaction, QCD, is
nonperturbative in the low-energy regime characteristic for
nuclear physics. The way out of this dilemma is the effective
field theory concept which recognizes different energy scales in
nature. Below the chiral symmetry breaking scale,
$\Lambda_\chi \approx 1$ GeV,
the appropriate degrees of freedom are pions and
nucleons interacting via a force that is governed by
the symmetries of QCD, particularly, (broken) chiral symmetry.

The derivation of the nuclear force from chiral effective field
theory was initiated by Weinberg \cite{Wei90} and pioneered
by Ord\'o\~nez \cite{OK92} and van Kolck \cite{ORK94,Kol99}.
Subsequently, many groups got involved in the subject
\cite{RR94,KBW97,KGW98,Kai99,KSW98,EGM98}.
As a result, efficient methods for deriving
the nuclear force from chiral Lagrangians have emerged.
Also, the quantitative nature of the chiral $NN$ potential
has improved \cite{EGM98}.
Nevertheless, even the currently `best' chiral $NN$ potentials
are too inaccurate to serve as a
reliable input for exact few-nucleon calculations or miscroscopic nuclear
many-body theory.

The time has come
to put the chiral approach to a real test in microscopic
nuclear structure physics. Conclusive results can, however, be produced
only with a 100\% quantitative $NN$ potential based upon chiral Lagrangians.
For this reason, we have embarked on a program to develop a $NN$ potential
that is based upon chiral effective field theory and reproduces the
$NN$ data with about that same quality as the
high-precision $NN$ potentials constructed in the 
1990's \cite{Sto94,WSS95,MSS96,Mac01}.

Starting point for the derivation of the $NN$ interaction is an
effective chiral $\pi N$ Lagrangian which is given by
a series of terms of increasing chiral dimension \cite{Fet00},
\begin{equation}
{\cal L}_{\pi N} 
=
{\cal L}_{\pi N}^{(1)} 
+
{\cal L}_{\pi N}^{(2)} 
+
{\cal L}_{\pi N}^{(3)} 
+ \ldots ,
\end{equation}
where the superscript refers to the number of derivatives or pion mass insertions
(chiral dimension)
and the ellipsis denotes terms of chiral order four or higher.

\begin{figure}
\vspace{-1.0cm}
\hspace{3.5cm}
\psfig{figure=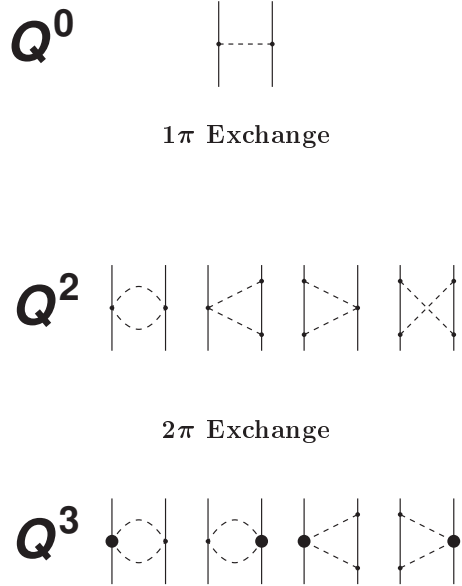,height=12cm}
\vspace{-2.5cm}
\caption{The most important irreducible one- and two-pion exchange contributions to the $NN$
interaction up to order $Q^3$. Vertices denoted by small dots are from
$\widehat{\cal L}^{(1)}_{\pi N}$,
while large dots refer to
$\widehat{\cal L}^{(2)}_{\pi N, \, \rm ct}$.}
\end{figure}

\begin{figure}
\vspace{-11.0cm}
\hspace{-2.2cm}\epsfig{figure=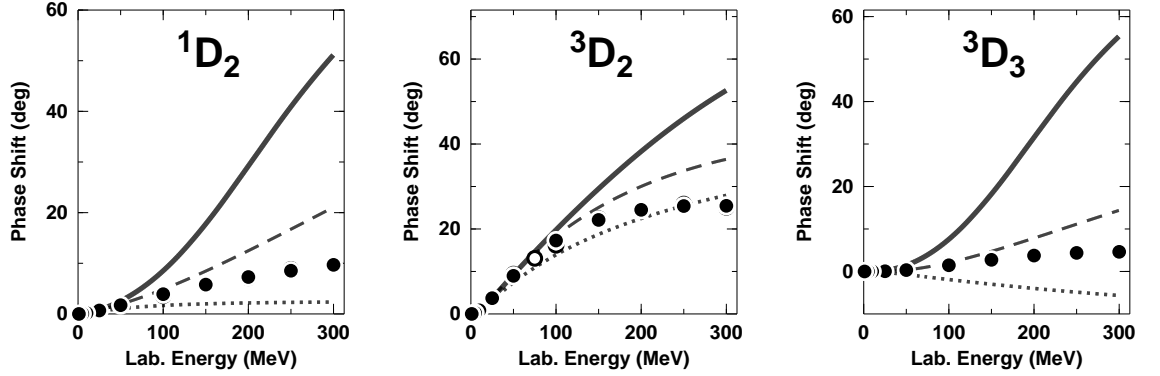,height=27cm}
\vspace{-10.5cm}
\caption{$D$-wave phase shifts of $NN$ scattering. The predictions by the chiral
model displayed in Fig.~1 are shown by the solid line and the ones by the
Bonn $\pi + 2\pi$ model~\protect\cite{MHE87} by the dashed curve. The dotted line is OPE.
Solid dots represent the Nijmegen multi-energy $np$ analysis~\protect\cite{Sto93}
and open circles the VPI/GWU analysis~\protect\cite{SM99}.}
\end{figure}

We will apply
the heavy baryon (HB) formulation of chiral perturbation theory \cite{BKM95}
in which the relativistic Lagrangian is subjected
to an expansion in terms of powers of $1/M_N$ (kind of a
nonrelativistic expansion), the lowest order of which is
\begin{eqnarray}
\widehat{\cal L}^{(1)}_{\pi N} & 
= & 
\bar{N} \left(
 i {D}_0 
 - \frac{g_A}{2} \; 
\vec \sigma \cdot \vec u
\right) N  
\\
 & 
\approx & 
\bar{N} \left[ i \partial_0 
- \frac{1}{4f_\pi^2} \;
\mbox{\boldmath $\tau$} \cdot 
 ( \mbox{\boldmath $\pi$}
\times
 \partial_0 \mbox{\boldmath $\pi$})
- \frac{g_A}{2f_\pi} \;
\mbox{\boldmath $\tau$} \cdot 
 ( \vec \sigma \cdot \vec \nabla )
\mbox{\boldmath $\pi$} \right] N + \ldots \, ,
\end{eqnarray}
where we use the notation of Ref.~\cite{BKM95}.
For the parameters that occur in the leading order Lagrangian,
we apply $M_N=938.919$ MeV, $m_\pi = 138.04$ MeV,
$f_\pi  =  92.4$ MeV, and
$g_A  =  g_{\pi NN} \; f_\pi/M_N = 1.29$,
which is equivalent to
$g_{\pi NN}^2/4\pi  =  13.67$.

The HB projected Lagrangian at order two is most conveniently broken up
into two pieces,
\begin{equation}
\widehat{\cal L}^{(2)}_{\pi N} \, = \,
\widehat{\cal L}^{(2)}_{\pi N, \, \rm fix} \, + \,
\widehat{\cal L}^{(2)}_{\pi N, \, \rm ct} \, ,
\end{equation}
with
\begin{equation}
\widehat{\cal L}^{(2)}_{\pi N, \, \rm fix}  =  
\bar{N} \left[
\frac{1}{2M_N}\: \vec D \cdot \vec D
 + i\, \frac{g_A}{4M_N}\: \{\vec \sigma \cdot \vec D, u_0\}
\right] N
\label{eq_L2fix}
\end{equation}
and
\begin{eqnarray}
\widehat{\cal L}^{(2)}_{\pi N, \, \rm ct}
& = & 
 \bar{N} \left[
2\,
c_1
\, m_\pi^2\, (U+U^\dagger)
\, + \, \left( 
c_2
- \frac{g_A^2}{8M_N}\right) u_0^2
 \, + \,
c_3
\, u_\mu  u^\mu
+ \, \frac{i}{2} \left( 
c_4
+ \frac{1}{4M_N} \right) 
  \vec \sigma \cdot ( \vec u \times \vec u)
 \right] N \, .
\end{eqnarray}
Note that
$\widehat{\cal L}^{(2)}_{\pi N, \, \rm fix}$
is created entirely from the HB expansion of the relativistic
${\cal L}^{(1)}_{\pi N}$ and thus has no free parameters (``fixed''),
while
$\widehat{\cal L}^{(2)}_{\pi N, \, \rm ct}$
is dominated by $\pi N$ contact terms proportional to the
$c_i$ parameters, besides some small $1/M_N$ corrections.
The parameters $c_i$ are known as low-enery constants (LECs)
and must be determined empirically from fits to $\pi N$ data.
We use the values determined by
B\"uttiker and Mei\ss ner \cite{BM00}
which are (in units of GeV$^{-1}$)
$c_1 = -0.81$,
$c_3 = -4.70$, and
$c_4 = 3.40$
($c_2$ will not be needed).

The $\pi N$ Lagrangian is the crucial ingredient for the evaluation of
the pion-exchange contributions to the $NN$ interaction.
Since we are dealing here with a low-energy
effective theory, it is appropriate to analyze the contributions
in terms of powers of small momenta: $(Q/\Lambda_\chi)^\nu$,
where $Q$ is a generic momentum or a pion mass and $\Lambda_\chi \approx 1$ GeV
is the chiral symmetry breaking scale.
This procedure has become known as power counting.
For the pion-exchange diagrams relevant to our problem, the power $\nu$ of a diagram
is determined by the simple formula
\begin{equation}
\nu = 2 \, l + \sum_j (d_j - 1) \, ,
\end{equation}
where $l$ denotes the number of loops in the diagram,
$d_j$ the number of derivatives involved in vertex $j$, and the sum
runs over all the vertices of the diagram.

The most important irreducible one-pion exchange (OPE) and two-pion exchange (TPE)
contributions to the $NN$ interaction
up to order $Q^3$ are shown in Fig.~1; they have been evaluated by
Kaiser {\it et al.} \cite{KBW97} using covariant perturbation theory and
dimensional regularization.
One- and two-pion exchanges are known to describe $NN$ scattering
in peripheral partial waves. In $G$ and higher partial waves 
(orbital angular momentum $L\geq 4$), there is good agreement
between the chiral and conventional~\cite{MHE87}
 $2\pi$ model as well as the empirical phase
shifts~\cite{Sto93,SM99}. 
The agreement deteriorates when proceeding to lower $L$. While in $F$ waves
the agreement between the chiral model and the empirical phase shifts is
still fair, substantial discrepancies emerge in $D$ waves, Fig.~2,
where the chiral $2\pi$ exchange is far too
attractive---a fact that has been noticed before \cite{KBW97,KGW98}.

To control the $D$ (and lower) partial waves, we need (repulsive) short-range contributions.
In the conventional meson model~\cite{MHE87}, these are created by the exchange of
heavy mesons (notably, the $\omega$ meson).
In chiral perturbation theory ($\chi$PT), heavy mesons have no place
and the short-range force is parametrized in terms of contact potentials,
which are organized by powers of $Q$.
If $Q$ is, e.~g., a momentum transfer, i.~e., $\vec Q = {\vec p}~' - \vec p$,
where $\vec p$ and ${\vec p}~'$ are the CM nucleon momenta before and after
scattering, respectively, and $\theta$ is the scattering angle, then,
for even $\nu$,
\begin{equation}
{\vec Q}^\nu \sim (\cos \theta)^m 
\hspace*{.5cm}
\mbox{with}
\hspace*{.5cm}
m\leq \frac{\nu}{2} 
\, .
\end{equation}
Partial-wave decomposition for orbital-angular momentum $L$
yields,
\begin{equation}
\int_{-1}^{+1} {\vec Q}^\nu P_L(\cos \theta) d\cos \theta \neq 0
\hspace*{.5cm}
\mbox{for}
\hspace*{.5cm}
L\leq \frac{\nu}{2} \, ,
\end{equation}
where $P_L$ is a Legendre polynominal.
The conclusion is that for non-vanishing contributions in $D$ waves ($L=2$),
$\nu=4$ is required.
{\it This one important message that we like to convey in this letter.}
Based upon invariance considerations, there are a total of 24 contact terms
up to order $Q^4$, which we all include in our model.
The parameters of these terms have to be natural,
but are otherwise unconstrained and, thus, represent essentially free
parameters.

\begin{figure}[t]
\vspace{-1.5cm}
\hspace{-2.3cm}
\psfig{figure=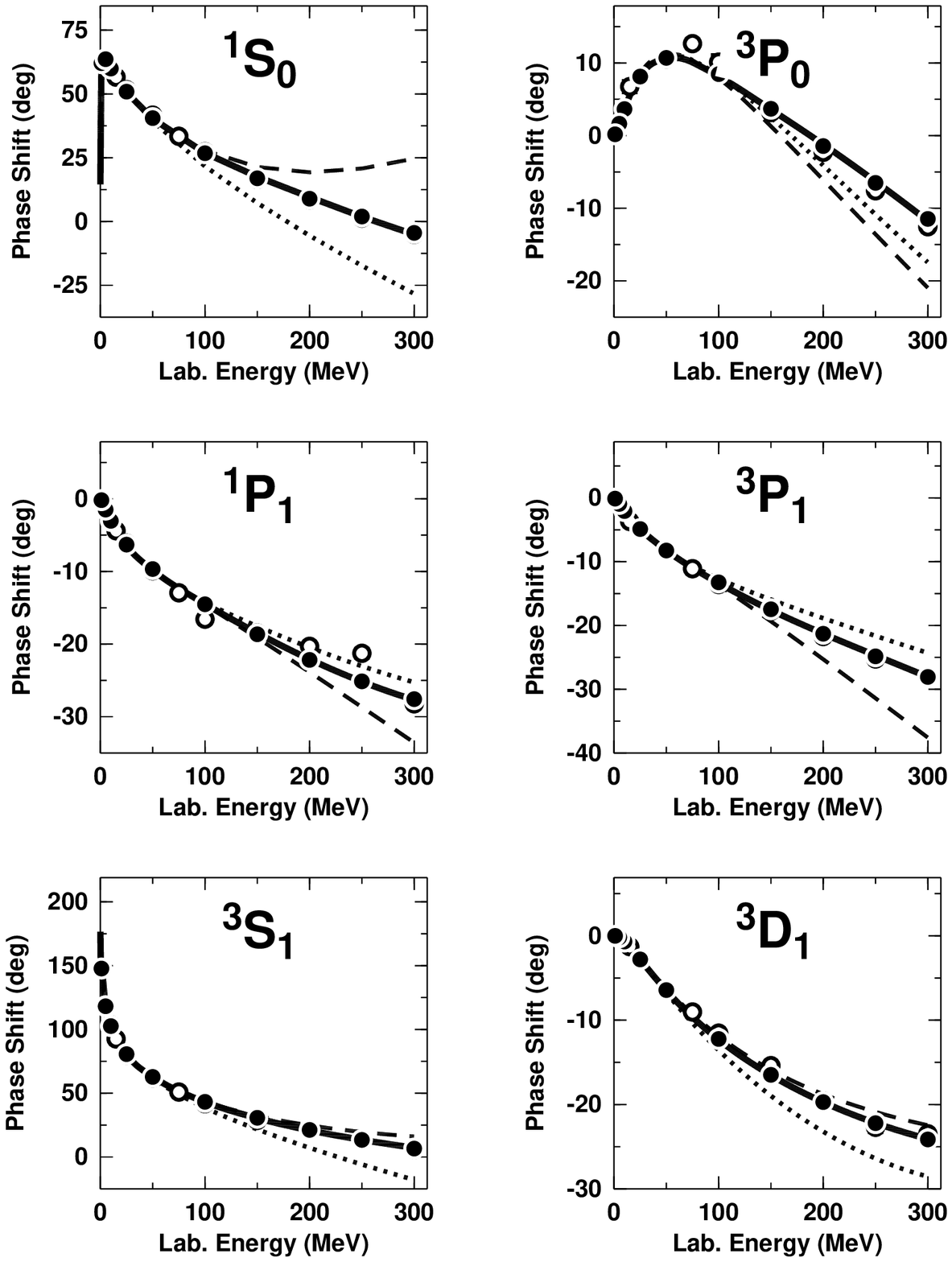,height=14.0cm}
\hspace{-3.2cm}
\psfig{figure=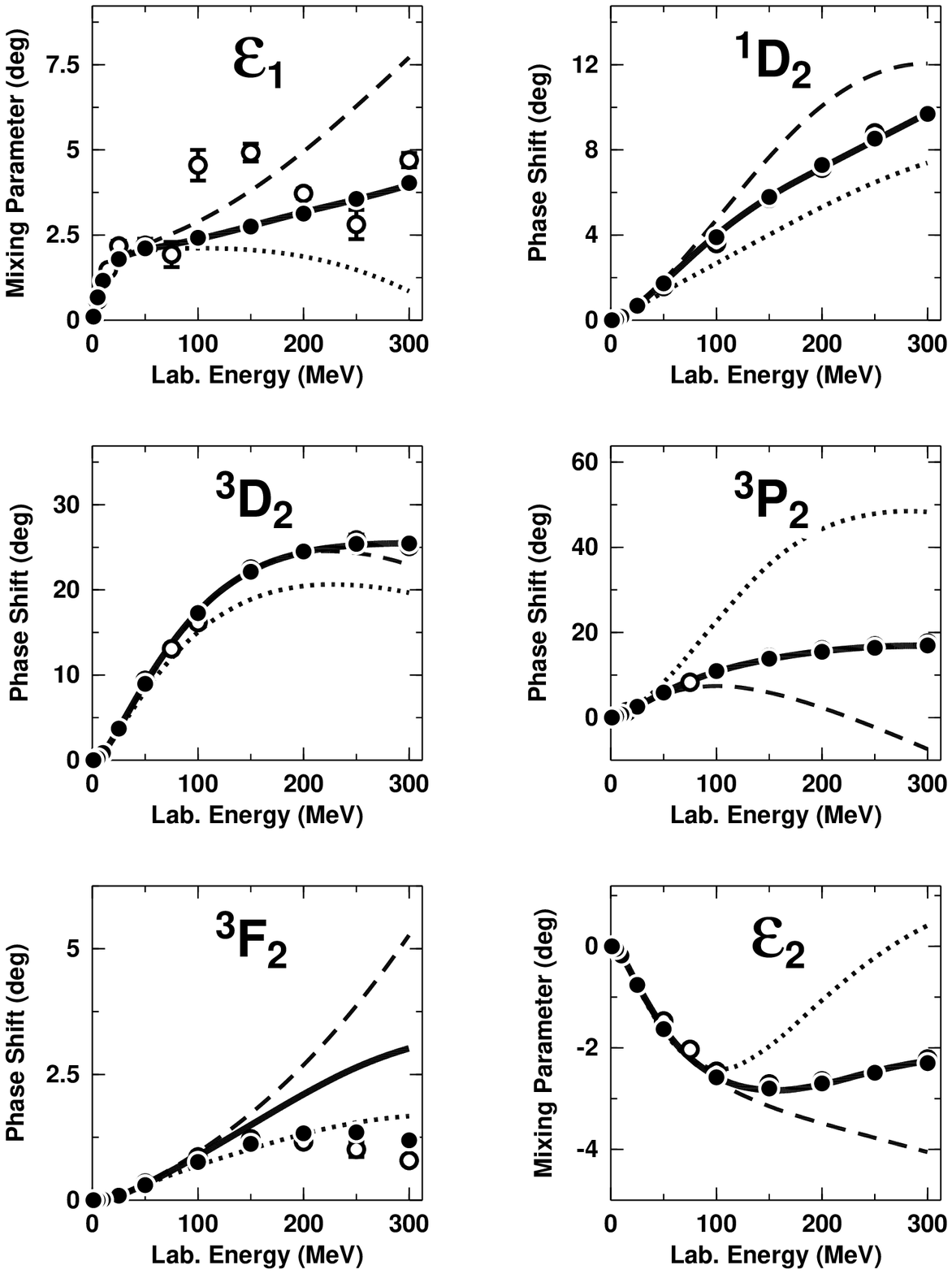,height=14.0cm}
\vspace{-2.2cm}
\caption{Phase shifts for $J\leq 2$.
The solid line is the result from our chiral $NN$ potential,
while the dotted and dashed lines are the predictions by two
chiral models developed by Epelbaum {\it et al.} \protect\cite{EGM98}
(NLO and NNLO, respectively).
Solid dots and open circles represent phase shift analyses explained
in the caption of Fig.~2.}
\end{figure}

To describe $NN$ scattering, we
start from
the Bethe-Salpeter (BS) equation \cite{SB51}
which reads in operator notation
\begin{equation}
{\cal T=V+V\,G\,T}
\end{equation}
with ${\cal T}$ the invariant amplitude for the two-nucleon scattering process,
${\cal V}$ the sum of all connected two-particle irreducible diagrams, and
${\cal G}$ the relativistic two-nucleon propagator.
The BS equation
is equivalent to a set of two equations:
\begin{eqnarray}
{\cal T}&=&\bar{V}+\bar{V}g\; {\cal T}
\label{eq_bbs}
\\
\bar{V}&=&{\cal V + V\;(G}-g)\bar{V}
\\
&\approx&
{\cal V + V_{\rm OPE}\;(G}-g){\cal V_{\rm OPE}}
\label{eq_box}
\end{eqnarray}
where the last line states the approximation we are using,
exhibiting the way we treat the $2\pi$ box diagram.
This treatment avoids double counting when $\bar V$ is
iterated in the scattering equation and is also
consistent with the calculations of Ref. \cite{KBW97}.
For the relativistic three-dimensional propagator $g$, we choose
the one proposed by
Blankenbecler and Sugar \cite{BS66}
(BbS)
which has the great
practical advantage that the OPE (and the entire potential)
becomes energy-independent.
Thus, we do not need the rather elaborate formalism of unitary
transformations \cite{EGM98} to generate energy-independence of the potential.

Our full chiral $NN$ potential $\bar V$ is defined by
\begin{equation}
\bar{V}({\vec p}~',{\vec p}) \equiv
\left\{ \begin{array}{c}
\mbox{ sum of irreducible}\\
\mbox{\boldmath $\pi + 2\pi$ contributions}
\end{array} \right\} 
+ \mbox{ contacts} \, ,
\label{eq_pot1}
\end{equation}
where the first term on the r.h.s.\ is given by
Eq.~(\ref{eq_box}) with $\cal V$ containing essentially the
diagrams of Fig.~1.
This potential satisfies the relativistic
BbS equation, Eq.~(\ref{eq_bbs}).
If we define now,
\begin{equation}
{V}({\vec p}~',{\vec p})
\equiv \sqrt{\frac{M_N}{E_{p'}}}\:  
\bar{V}({\vec p}~',{\vec p})\:
\sqrt{\frac{M_N}{E_{p}}}
\approx \left(1
-\frac{p'^2+p^2}{4M_N^2}
\right)
\bar{V}({\vec p}~',{\vec p}) 
\label{eq_pot2}
\end{equation}
with $E_p\equiv \sqrt{M_N^2 + {\vec p}^2}$,
then $V$ satisfies the usual, nonrelativistic
Lippmann-Schwinger (LS) equation.

Iteration of $V$ in the LS equation
requires cutting $V$ off for high momenta to avoid infinities.
Therefore, we regularize $V$ in the following way:
\begin{eqnarray}
V(\vec{ p}~',{\vec p})& 
\longmapsto&
V(\vec{ p}~',{\vec p})
\;\mbox{\boldmath $e$}^{-(p'/\Lambda)^{2n}}
\;\mbox{\boldmath $e$}^{-(p/\Lambda)^{2n}}
\\
&&
\approx
V(\vec{ p}~',{\vec p})
\left\{1-\left[\left(\frac{p'}{\Lambda}\right)^{2n}
+\left(\frac{p}{\Lambda}\right)^{2n}\right]+ \ldots \right\} \, ,
\end{eqnarray}
where the last equation is to indicate that
the exponential cutoff does not affect the order to which we are
calculating, but introduces contributions beyond that order.
For the contact terms, we use partial wave dependent cutoff
parameters $\Lambda \approx 0.4 - 0.5$ GeV which brings the total
number of parameters in our chiral $NN$ model up to 46.
At first glance, this may appear to be a large number.
Note, however, that the Nijmegen phase shift analysis~\cite{Sto93}
and the high-precision potentials~\cite{Sto94,WSS95,MSS96,Mac01}
developed in the 1990's carry between 40 and 50 parameters.
Thus, the number of parameters needed for
a quantitative chiral $NN$ model is just about the same as for meson models.
Since the chiral model has less predictive power than the meson model
this should not be unexpected.

\begin{table}
\caption{Two- and three-nucleon low-energy data.}
\footnotesize
\begin{tabular}{llllll}
 & Idaho-A$^a$ & Idaho-B$^a$ & CD-Bonn\cite{Mac01} & AV18\cite{WSS95} & Empirical$^b$ \\
\hline
\hline
\multicolumn{2}{l}{\bf Low-energy $np$ scattering} \\
$^1S_0$ scattering length (fm) &-23.75&-23.75&-23.74&-23.73&-23.74(2)\\
$^1S_0$ effective range  (fm) &2.70&2.70&2.67&2.70&2.77(5)\\
$^3S_1$ scattering length (fm) &5.417&5.417&5.420&5.419&5.419(7)\\
$^3S_1$ effective range (fm) &1.750&1.750&1.751&1.753&1.753(8)\\
\hline
{\bf Deuteron properties}\\
Binding energy (MeV) &
2.224575& 2.224575&
 2.224575 & 2.224575 & 2.224575(9) \\
Asympt.\ $S$ state (fm$^{-1/2}$) &
0.8846 &0.8846 
& 0.8846& 0.8850 & 0.8846(9) \\
Asympt. $D/S$ state         & 
0.0256 & 0.0255 &
0.0256& 0.0250&0.0256(4)\\
Deuteron radius (fm)   &
1.9756$^c$ &
1.9758$^c$ &
 1.970$^c$ &
 1.971$^c$ &
 1.9754(9)$^d$ \\
Quadrupole moment (fm$^2$) &
0.281$^e$ &
0.284$^e$ &
 0.280$^e$ & 
0.280$^e$ &
 0.2859(3) \\
$D$-state probability (\%)    & 
4.17 &
4.94 &
4.85 & 5.76  \\
\hline
{\bf Triton binding} (MeV) &8.14&8.02&8.00&7.62&8.48\\
\end{tabular}
\indent
$^a$Chiral NN potential of the present work.\\
$^b$For references concerning the empirical data, 
see Tables XIV and XVIII of Ref.\cite{Mac01}.\\
$^c$With meson-exchange current (MEC) and relativistic corrections~\cite{FMS97}.\\
$^d$Reference\cite{Hub98}.\\
$^e$Including MEC and relativistic corrections 
in the amount of 0.010 fm$^2$~\cite{Hen95}.
\end{table}

In Fig.~3, we show the phase shifts of neutron-proton ($np$) scattering
for lab.\ energies below 300 MeV and partial waves with $J\leq 2$.
The solid line represents the result from the chiral $NN$ potential developed in
the present work. {\it The reproduction of the empirical phase shifts by our model
is excellent.} For comparison, we also show the phase shift predictions
by two chiral models recently developed by
Epelbaum {\it et al.} \cite{EGM98} (dotted and dashed curves in Fig.~3).
Also the effective range parameters in $S$ waves agree accurately with the 
empirical values, as well as the deuteron parameters (see Table I).
We note that our present chiral potential is charge-independent and adjusted
to the $np$ data.

Due to the very quantitative nature of this new chiral $NN$ potential,
it represents a reliable and promising starting point
for exact few-body calculations and
microscopic nuclear many-body theory.

A crucial finding of our investigation is that contact terms of order
four are required for a quantitative $NN$ model.
The basic ideas of $\chi$PT may then suggest that---for reasons of
consistency---the chiral $2\pi$ exchange contribution should also
be included up to order four. Therefore,
an implication of the present work is that the chiral
$2\pi$ exchange at order four~\cite{Kai01} will be important for further 
chiral $NN$ potential development.

We gratefully acknowledge useful discussions with B. van Kolck,
E. Epelbaum, W. Gl\"ockle, N. Kaiser, U. Mei\ss ner, and M. Robilotta.
This work was supported in part by the U.S. National Science
Foundation under Grant No.~PHY-0099444 and by the Ram\'on Areces
Foundation (Spain).


\begin{thebibliography}{99}

\bibitem{Wei90} S. Weinberg, {\it Phys.\ Lett.\/} B {\bf 251}, 288 (1990);
{\it Nucl.\ Phys.\/} {\bf B363}, 3 (1991).

\bibitem{OK92}
C. Ord\'o\~nez
and U. van Kolck,
{\it Phys.\ Lett.\/} B {\bf 291}, 459 (1992).

\bibitem{ORK94}
C. Ord\'o\~nez,
L. Ray, and U. van Kolck,
{\it Phys.\ Rev.\ Lett.\/} {\bf 72}, 1982 (1994);
{\it Phys.\ Rev.\/} C {\bf 53}, 2086 (1996).

\bibitem{Kol99} U. van Kolck, {\it Prog.\ Part.\ Nucl.\ Phys.\/} {\bf 43}, 337 (1999).

\bibitem{RR94}
C. A. da Rocha and M. R. Robilotta, {\it Phys.\ Rev.\/} C {\bf 49}, 1818 (1994);
{\it ibid.} {\bf 52}, 531 (1995);
J.-L. Ballot {\it et al.\/},
{\it ibid.\/} C {\bf 57}, 1574 (1998).

\bibitem{KBW97} N. Kaiser, R. Brockmann, and W. Weise,
{\it Nucl.\ Phys.\/} {\bf A625}, 758 (1997).

\bibitem{KGW98} N. Kaiser {\it et al.\/},
{\it Nucl.\ Phys.\/} {\bf A637}, 395 (1998).

\bibitem{Kai99} N. Kaiser, {\it Phys.\ Rev.\/} C
{\bf 61}, 014003 (1999);
{\it ibid.\/} {\bf 62}, 024001 (2000);
{\it ibid.\/} {\bf 63}, 044010 (2001).

\bibitem{KSW98} D. B. Kaplan, M. J. Savage, and M. B. Wise,
{\it Nucl.\ Phys.\/} {\bf B534}, 329 (1998).

\bibitem{EGM98} E. Epelbaum, W. Gl\"ockle, and U.-G. Mei\ss ner,
{\it Nucl.\ Phys.\/} {\bf A637}, 107 (1998); {\it ibid.\/}
{\bf A671}, 295 (2000).

\bibitem{Sto94} V.\ G.\ J.\ Stoks {\it et al.\/},
{\it Phys.\ Rev.\/} C {\bf 49}, 2950 (1994).

\bibitem{WSS95} R.\ B.\ Wiringa {\it et al.\/},
{\it Phys.\ Rev.\/} C {\bf 51}, 38 (1995).

\bibitem{MSS96}
R. Machleidt, F. Sammarruca, and Y. Song, 
{\it Phys.\ Rev.\/} C {\bf 53}, 1483 (1996).

\bibitem{Mac01} R. Machleidt,
{\it Phys.\ Rev.\/} C {\bf 63}, 024001 (2001).

\bibitem{Fet00} N. Fettes, U.-G. Mei\ss ner, M. Moj\v{z}i\v{s}, and S. Steininger,
{\it Ann.\ Phys.\ (N.Y.)\/}
{\bf 283}, 273 (2000);
{\it ibid.\/} {\bf 288}, 249 (2001).

\bibitem{BKM95} V. Bernard {\it et al.\/},
{\it Int.\ J.\ Mod.\ Phys.\/} E {\bf 4}, 193 (1995).

\bibitem{BM00} P. B\"{u}ttiker and U.-G. Mei\ss ner,
{\it Nucl.\ Phys.\/} {\bf A668}, 97 (2000).

\bibitem{MHE87} R. Machleidt, K. Holinde, and Ch. Elster,
{\it Phys.\ Rep.\/} {\bf 149}, 1 (1987).

\bibitem{Sto93} V.\ G.\ J.\ Stoks {\it et al.\/},
{\it Phys.\ Rev.\/} C {\bf 48}, 792 (1993).

\bibitem{SM99}
R. A. Arndt {\it et al.\/}, SAID, Solution SM99 (Summer 1999).

\bibitem{SB51} E. E. Salpeter and H. A. Bethe,
{\it Phys.\ Rev.\/} {\bf 84}, 1232 (1951).

\bibitem{BS66} R. Blankenbecler and R. Sugar,
{\it Phys.\ Rev.\/} {\bf 142}, 1051 (1966).

\bibitem{FMS97} J. L. Friar {\it et al.\/},
{\it Phys.\ Rev.\/} A {\bf 56}, 4579 (1997).

\bibitem{Hub98} A. Huber {\it et al.}, 
{\it Phys.\ Rev.\ Lett.\/} {\bf 80}, 468 (1998).

\bibitem{Hen95} J. Adam and H. Henning, private communication.

\bibitem{Kai01}
N. Kaiser, {\it Chiral $2\pi$-exchange NN potential: Two-loop
contributions}, {\tt nucl-th/0107064}.

\end{thebibliography}
\end{document}